\documentclass{article}

\usepackage{arxiv}

\usepackage[utf8]{inputenc} 
\usepackage[T1]{fontenc}    
\usepackage{hyperref}       
\usepackage{url}            

\usepackage{booktabs}       
\usepackage{amsfonts}       
\usepackage{nicefrac}       
\usepackage{microtype}      

\usepackage{times}
\usepackage{soul}
\usepackage{url}
\usepackage{graphicx}
\usepackage{amsmath}
\usepackage{booktabs}
\usepackage[table,xcdraw]{xcolor}
\usepackage{algorithm}
\usepackage{amssymb}
\usepackage[noend]{algpseudocode}
\usepackage{multirow}
\usepackage{xcolor}
\usepackage{lineno,hyperref}
\usepackage{geometry}
\modulolinenumbers[5]

\title{
Fake News Data Collection and Classification: \\
Iterative Query Selection for Opaque Search Engines with Pseudo Relevance Feedback}

\author{Aviad Elyashar, Maor Reuben, and Rami Puzis \\
Telekom Innovation Laboratories and Department of Software and Information Systems Engineering,\\
Ben-Gurion University of the Negev, Beer-Sheva, Israel \\
\{aviade,maorreu\}@post.bgu.ac.il, puzis@bgu.ac.il}

\begin{document}
\maketitle

\begin{abstract}
Retrieving information from an online search engine, is the first and most important step in many data mining tasks.   
Most of the search engines currently available on the web, including all social media platforms, are black-boxes (a.k.a opaque) supporting short keyword queries. 
In these settings, retrieving all posts and comments discussing a particular news item automatically and at large scales is a challenging task. 
In this paper, we propose a method for generating short keyword queries given a prototype document.
The proposed \emph{iterative query selection} algorithm (IQS) interacts with the opaque search engine to iteratively improve the query. 
It is evaluated on the Twitter TREC Microblog 2012 and TREC-COVID 2019 datasets showing superior performance compared to state-of-the-art.
IQS is applied to automatically collect a large-scale fake news dataset of about 70K true and fake news items. 
The dataset, publicly available for research, includes more than 22M accounts and 61M tweets in Twitter approved format. 
We demonstrate the usefulness of the dataset for fake news detection task achieving state-of-the-art performance.
\end{abstract}


\section{Introduction}
Every day, millions of people search for information online~\cite{christopher2008introduction}. 
Market researchers search for products related to their product or business~\cite{yao2012product}. 
Researchers reviewing the academic literature search for works related to their current article~\cite{bethard2010should}.
Posts and comments that discuss a news item are retrieved from online social media (OSM) for fake news detection~\cite{zhou2015real}.
There are many similar cases when additional information related to a specific document is required. 
We will refer to such a document as a \emph{prototype}.

There are multiple methods for retrieving a set of documents that are similar to a given prototype from a corpus. 
Most of these methods represent documents as vectors and calculate the similarity between the prototype and other documents.   
The basic methods are based on TF-IDF (term frequency-inverse document frequency) and BM25 which treats each document as a bag-of-words~\cite{alvarez2017review}.
Advanced methods, such as Doc2Vec~\cite{le2014distributed}, Skip-thoughts~\cite{kiros2015skip}, Sent2Vec~\cite{pagliardini2018unsupervised}, and others, use neural networks to represent documents as low-dimensional vectors~\cite{alvarez2017review}.
Once vector representations of documents are available, retrieving the documents that are most similar to the prototype, i.e., closest to it in the embedding space, is straightforward. 

Retrieval methods that are based on document similarity assume access to the corpus being searched. 
This assumption is valid for transparent search engines, where the repository and the algorithms are known to the user.
However, all the popular search engines, including general-purpose search like Google, or platform-specific, such as Twitter search, are opaque providing very little information about their repositories and algorithms~\cite{urgen1996case}.
Other than Google's image search, current search engines do not provide document-based search services. 
Therefore, we usually resort to the short keyword queries that are a mainstay of anyone using today's search engines~\cite{chirita2007personalized}.

Due to the ambiguity of short keyword queries, they often do not reflect the original intention of the query writer~\cite{cronen2002predicting}.
For example, the keyword ``apple'' may refer to the fruit or to the technology company.%
\footnote{Google and some other engines use the search context to resolve ambiguity and retrieve documents that are most relevant for a specific user or case~\cite{finkelstein2001placing}.
Other engines, such as Twitter, do not use contextual information and retrieve all results exactly matching the specified keywords~\cite{searchtweets}.
Personalization and context aware information retrieval are out of the scope of this paper. } 
In this paper, we focus on the problem of \emph{retrieving documents that are most similar to a given prototype document from an opaque search engine supporting short keyword queries}, such as Twitter.
It is possible to manually formulate a search query from the document's content~\cite{zhou2015real}. 
But of course, manual query selection does not scale. 
One can use the prototype document's title to generate queries~\cite{monti2019fake},
or search for its URL if the prototype is a web page~\cite{vosoughi2018spread}.
However, these approaches miss many relevant results. 
We further discuss the pros and cons of existing query selection methods in Section~\ref{sec:related_work}. 

Therefore, in this paper, we suggest a novel iterative approach that selects queries that maximize the number of retrieved relevant results, using limited interaction with an opaque search engine. 
This approach consists of two components: 
the \emph{iterative query selection (IQS)} algorithm and the \emph{word's mover distance (WMD)} measure. 
The \emph{IQS} is a hill climbing algorithm that iteratively optimizes short keyword queries given a prototype document.
The \emph{WMD} is used as pseudo relevance feedback, by ranking the results of incumbent queries generated by the IQS algorithm according to their relevance to the prototype document.
The details of the proposed method are discussed in Section~\ref{sec:iterative_keyword_optimization}.
In the absence of a prototype document, incumbent results are compared to a set of relevant results according to user relevance feedback. 
We evaluated the proposed methods on TREC 2012 Microblog benchmark and TREC-COVID 2019 datasets assuming relevance feedback (see Section~\ref{sec:methods_evaluation})
In addition, we utilized the proposed IQS algorithm to retrieve a large-scale fake news dataset from Twitter to be used toward training fake news classifiers (see  Section~\ref{sec:fake_news_classification}).

The contributions of this paper are:
\begin{itemize}
\item automated mechanism for optimizing short keyword queries termed \emph{iterative query selection (IQS)} (see Section~\ref{sec:hill_climbing}). 
The IQS outperforms existing opaque relevance feedback search on Twitter TREC Microblog 2012 and on TREC-COVID 2019 datasets (see Section~\ref{sec:hill_climbing_evaluation}). 
\item a large-scale \emph{Fake News} dataset:\footnote{The dataset is publicly available as a collection of Twitter IDs in the following link: \url{https://bit.ly/2vd58u6}}
70K news items discussed by 20M twitter users in 61M tweets (see Section~\ref{sec:iqs-fake-news-collection}).
\item the quality of the dataset is demonstrated through the application of fake news detection algorithms on the collected data achieving  AUC\footnote{Area Under the Receiver Operating Characteristic Curve} of 0.92 and accuracy of 0.86 (see Section~\ref{sec:fake_news_results}).
\end{itemize}


The rest of the paper is organized as follows: 
In Section~\ref{sec:related_work}, we review previous approaches for query selection, result diversification, document similarity, and fake news collection.
In Section~\ref{sec:iterative_keyword_optimization}, we present the proposed iterative query selection algorithm for optimizing short keyword queries sent to an opaque search engine.
In Section~\ref{sec:methods_evaluation}, we present the datasets used for evaluating the solution proposed, as well as discuss the results obtained.
Section~\ref{sec:ethical_considerations} discusses ethical considerations, and
we conclude the paper in Section~\ref{sec:conclusion} with our plans for future work.


\section{Related Work}
\label{sec:related_work}
This paper describes a new approach for selecting the best queries engaging with opaque search engines.
In the next sections, we elaborate on the methods that are associated with the query selection.
Next, for the demonstration of fake news detection use cases, we provide the necessary background for this domain.

\subsection{Query Selection for Transparent Search Engines}
Query selection is the task of selecting the most suitable queries for extracting relevant documents from web search engines~\cite{wu2006query}.
In most cases, selecting these queries requires reformulation or expansion of an initial query.
Several methods suggest analyzing the underlying corpus of the given search engine and used this valuable information for expanding the queries.
Dwaipayan et al.~\cite{roy2016using} selected the most similar terms to a given query for expansion based on word embedding trained on the corpus.
Their idea was to choose terms that yielded the highest probability of being related to the current query.
Kuzi et al.~\cite{kuzi2016query} like Dwaipayan et al. used word embedding trained on the corpus to select expansion terms and suggested centroid- and fusion-based terms scoring methods to select them.
Xu et al.~\cite{xu2018improving} selected candidate terms for expansion based on context features, such as TF-IDF and co-occurrence of the query terms.
Afterward, they used the learned term-ranking models to rank the candidate terms.
Pang and Du~\cite{pang2019query} utilized click-through data of old queries for query reformulation.
They first construct a click network that consists of queries and documents as nodes.
Then they calculate the conditional probability of each term from the neighbor queries to be in the input query.
Finally, they use the top terms to expand short queries and the tail terms to reduce long queries.

All these approaches require knowledge about the underlying corpus of the search engine and therefore are not suitable for the case of opaque search engines. 
Also, their starting point requires an initial query. 
Thus, it is not possible to use a prototype document in these approaches.   

\subsection{Query Selection for Opaque Search Engines}

In opaque search engines, we lack knowledge about the underlying search method, corpus, or query selection method (if there is one).
Thus, to optimize a query, we require an external query selection method.
Such methods expand and reformulate an initial query using interactions with the search engine.
Li et al.~\cite{li2014req} presented ReQ-ReC (ReQuery-ReClassify) a double-loop active retrieval system.
The double-loop is a combination of an outer loop that is responsible for selecting new queries, and an inner loop that trains a document ranker using active learning.
The process is finished when there are no more documents labeled as relevant from the user, or the user is satisfied with the results.
ATR-Vis (Active Tweet Retrieval Visualization) is another retrieval system that was presented by Makki et al.~\cite{makki2018atr}. 
This system is interactive and exploratory tool that detects tweets that are related to a given debate. 
To decrease user involvement in the process, ATR-Vis proposes four strategies of active learning.
Ambiguous retrieval strategy sends tweets, that have a similar probability to relate to more than one debate, to labeling.
In the near-duplicates strategy, tweets that have similar text, are labeled the same.
The Leveraging hashtags strategy filters tweets containing hashtags that appear in multiple debates.
In the leveraging replies strategy, tweets that their replies are classified uniformly among all debates are sent for labeling. 
Zamani et al.~\cite{zamani2016pseudo} referred to the task of query expansion as a recommendation task.
First, they consider the query and the retrieved pseudo-relevant documents as users, and the terms as items.
Then they use non-negative matrix factorization to recommend terms for the given query.
Another approach for query reformulation was introduced by Al-Khateeb et al.~\cite{al2017query}, where the initial query can be reformulated using a genetic algorithm search.
The synonyms of the query terms are candidates for the reformulation, and the fitness function is based on the similarity between the query and the results.
Nogueira and Cho~\cite{nogueira2017task} presented a neural network architecture that reformulates a query.
The network receives the query terms and a given candidate term as input. Then, it predicts whether the candidate term is suitable for expanding the query.
Chy et al.~\cite{chy2019query} proposed a query expansion method that selects effective expansion terms using a random forest trained on term features.
The extracted features are grouped into five categories: lexical features, Twitter-specific features, temporal features, sentiment features, and embedding-based features. 
ALMIK is another active retrieval method that tries to achieve both high-precision and high-recall in collecting event-related tweets~\cite{zheng2019collecting}.
Similar to the ReQ-ReC method, ALMIK  contains a keyword expansion component, that improves the initial set of keywords iteratively, and an event-related tweet classifier that identifies related tweets.
To reduce annotation effort, the ALMIK tweet-related classifier is trained using a multiple-instance learning process.
This process assigns labels to bags of similar instances.


These approaches, except for ATR-Vis, require an initial query.
The drawback of the ATR-Vis is that it requires users to label the retrieved results (relevance feedback).
In contrast, our proposed method uses a pseudo-relevance feedback process that does not require user interaction and can be used on any search engine.

\subsection{Document Similarity}
\label{sec:doument_similarity}
Over the past years, various solutions have been suggested for estimating the semantic similarity between documents based on lexical matching, handcrafted patterns, syntactic parse trees, external sources of structured semantic knowledge, and distributional semantics.

We can divide document similarity measures into two groups: the supervised measures and the unsupervised measures.
The supervised measures require training to provide a similarity score for a pair of documents.
Kenter and de Rijke~\cite{kenter2015short} generate multiple types of meta-features from texts' word embedding to train a supervised learning classifier.
Later, they used the trained model for predicting the semantic similarity of new, unlabelled pairs of short texts.
Deep relevance matching model (DRMM)~\cite{guo2016deep} is a supervised model for determining the relevance of a document given a particular query.
The proposed model employed a joint deep architecture at the query term level that estimates the query document similarity. 
Mitra et al.~\cite{mitra2017learning} also suggested a supervised document ranking model composed of two separate deep neural networks, where the first network matches the query and the document using a local representation and the second network matches the query and the document using learned distributed representations. 
Next, the two networks are jointly trained as part of a single neural network. 
Mitra et al. showed that this combination performed better than either neural network individually on a web page ranking task, and significantly outperforms traditional baselines and other recently proposed models based on neural networks.

The unsupervised document similarity measures provide a similarity score for a pair of texts without the requirement of training.
The dual embedding space model (DESM) proposed by Nalisnick et al.~\cite{nalisnick2016improving} calculates the average cosine distance of each term in the query with the centroids of the documents using pre-trained word embeddings.
Another unsupervised document similarity measure is the word mover's distance (WMD) proposed by Kusner et al.~\cite{kusner2015word}.
It measures the dissimilarity between two documents as the minimal sum of distances that the word vectors of one document need to move towards the word vectors of another document.
In this paper, we use WMD and extend it to a collection of retrieved documents.

\subsection{Search Result Diversification}
\label{sec:search_result_diversification}
In many cases, queries for search engines can arguably be considered ambiguous to some extent. 
Therefore, in order to tackle query ambiguity, search result diversification approaches have recently been proposed to produce rankings for satisfying the multiple possible information needs underlying a query~\cite{drosou2010search}.
In most cases, the diversification of retrieved results implies a trade-off between having more relevant results that reflect the true intent of the user and having less redundancy~\cite{gollapudi2009axiomatic}. 
There are two prominent diversification approaches: implicit and explicit. 
The former approach implicitly assumes that similar documents will cover similar interpretations or aspects associated with the query and should hence be dismissed. 
In particular, an implicit representation of aspects relies on document features such as the terms contained in the retrieved documents~\cite{carbonell1998use}, the clicks they received~\cite{slivkins2010learning}, their topic models~\cite{carterette2009probabilistic}, or clusters~\cite{he2011result}.
The latter, explicit approach, allows a broad topic associated with an ambiguous query to be decomposed into its constituent sub-topics.
Therefore, we can explicitly search for different aspects of the query for producing a diverse ranking of results. 
%
In most of the cases, explicit approaches rely on features derived from the query as candidate aspects, such as different query categories~\cite{agrawal2009diversifying} or query reformulations~\cite{santos2010exploiting}.

In this paper, we diversify the returned documents using two query expansion methods: adding synonyms based on WordNet or Adding the $k$ closest words in the embedding space for each candidate keyword in the query.

\subsection{Fake News Data Collection Methods}
Fake news is a long-lasting problem that has drawn significant attention in recent years.
It has been widely spread within the online social media (OSM)~\cite{silverman2016analysis}.
Since the detection of fake news is very challenging, many researchers suggested different approaches to confronting this issue.
Many of them were based on natural language processing~\cite{zhou2019fakenlp}, investigating the diffusion of news~\cite{vosoughi2018spread}, etc. 
Also, a few papers have attempted to detect fake news solely using social context features~\cite{shu2017fake}.

In order to train supervised classifiers for fake news detection, a ground truth dataset containing labeled news items is required. 
Such news items can be collected from fact checking websites, such as Snopes,\footnote{https://www.snopes.com/} PolitiFact,\footnote{https://www.politifact.com/} FactCheck,\footnote{https://www.factcheck.org/} and others~\cite{vosoughi2018spread,wang2017liar}.  

There are two commonly-used methods for collecting relevant posts associated with a given claim:
The first method is to retrieve posts based on the sources that distributed the claims.
For example, Monti et al.~\cite{monti2019fake} used the source's headlines that exist in fact-checking websites to collect tweets.
Vosoughi et al.~\cite{vosoughi2018spread} investigated the diffusion of news, based on collected tweets that contained links to the given claims. 
However, collecting tweets based on sources may be incomplete since many posts are associated with the given claim, but do not contain a link to the claim's source. 
Moreover, URL shortening, quotation, and cross reference common in the press, as well as among bloggers lead to a situation where tweets mentioning the same news contain links to different sources.  
Therefore, collecting tweets only based on links will result in a subset of the tweets relevant to a claim. 
In addition, also the use of the source's headlines does not always reflect well the claim's content (e.g., in the case of clickbait) which can lead to irrelevant results.
These drawbacks limit the ability to collect a quality dataset that contains enough relevant data for accurate classification.

The second method that people use to collect relevant posts is through the use of manual query selection.
For example, Zhou et al.\cite{zhou2015real} demonstrated a real-time news certification system on Sina Weibo\footnote{https://www.weibo.com/} using queries provided by the user to gather related posts. 
Then, they built an ensemble model that combined user-based, propagation-based, and content-based features and evaluated the proposed model on a small dataset of 146 claims.
Jin et al.~\cite{jin2017multimodal} and Wang et al.~\cite{wang2018eann} developed neural network-based methods for fake news detection and to evaluate their proposed methods they both used two small datasets from Sina Weibo (40k tweets) and Twitter (15k~ tweets on 52 rumor-related events).
Those datasets were created using manual query selection.
Selecting queries manually to a large collection of claims requires a lot of human effort and limits the amount of collected data.

Due to the limitations of both methods described above, it is clear that fake news detection based on the OSM can benefit from a tool that can automatically select accurate short keyword queries for a given claim (i.e., a news item).
In this study, we demonstrate the usefulness of the proposed \emph{iterative query selection (IQS)} method to retrieve a large-scale fake news dataset from Twitter automatically, as well as train fake news classifiers using social context features extracted from the tweets.  


\section{Iterative Query Selection with Word Mover's Distance Objective Function}
\label{sec:iterative_keyword_optimization}
In this paper, we propose a novel iterative approach for optimizing short keyword queries given a prototype document through interaction with an opaque search engine.
First, we describe the \emph{word mover's distance (WMD)}, a measure suggested by Kusner et al.~\cite{kusner2015word} that estimates the similarity of results to a given prototype document. 
This measure is calculated by summing the shortest distances between words in the given prototype document and words in the retrieved results (see Section~\ref{sec:mean_relevance_error}). 
The lower the WMD, the more relevant the retrieved results are.
Second, 
we outline the \emph{iterative query selection (IQS)} algorithm for finding queries that retrieve results with the lowest \emph{WMD} score (see Section~\ref{sec:hill_climbing}).

\subsection{Word Mover's Distance}
\label{sec:mean_relevance_error}
In this section, we describe the word mover's distance (WMD) measures and its aggregation for multiple documents the mean word mover's distance (MMD).

The WMD measure estimates the minimal distance between word vector representations of the words existing in the retrieved result and the prototype document. 
The intuition is that documents that are close in their semantic space, probably discuss the same topic.

Let \(d\) denote the prototype document and \(r\) denote a short document retrieved using a search engine. 
Let $w$ be a word vector representation that is calculated using a pre-traind word embedding method such as GloVe~\cite{pennington2014glove}, Word2Vec~\cite{mikolov2013efficient}, FastText~\cite{bojanowski2016enriching}, etc.

We can use any word embedding method, where words with a similar meaning are embedded close to each other.  
Let \(dist(w_i,w_j)\) denotes the cosine distance between the vector representations of two words (\(w_i,w_j\in \mathbb{R}^n\)).
The cosine distance is defined as $1 - cosineSim(w_i, w_j)$.
Thus the cosine distance ranges between 0 to 2.

Let \(W_d=\{w_{d_1}, w_{d_2},\dots,w_{d_l}\}\) be the set of word vectors in \(d\) and
$W_r=\{w_{r_1}, w_{r_2},\dots, w_{r_k}\}$ be the set of word vectors in \(r\).
The \(W_d\) and \(W_r\) do not contain stop words. 
The distance between a word \(w\) and a document \(d\) is the minimal distance between the word \(w_i\) and all the words in \(W_d\) (see Equation~\ref{eq:dist}). 
\begin{equation}
\label{eq:dist}
 dist(w_{r_i}, d) = \min_{w_{d_j}\in W_d} \{dist(w_{r_i}, w_{d_j})\}
\end{equation}


The distance between the word vectors of the word \(w_{r_i}\) and all the words in \(W_d\) reflects the semantic similarity of the word \(w_{r_i}\) to the prototype document.
The smaller the distance the higher semantic similarity.

Given a result document $r$, let word mover's distance ($WMD$) of $r$ with respect to the prototype document $d$ be the average distance of all words $w_{r_i} \in W_r$ to the document $d$ (see Equation~\ref{eq:re}): 
\begin{equation}
\label{eq:re}
    WMD(r, d) = \frac{1}{|W_r|}\sum_{w_{r_i}\in W_r} dist(w_{r_i},W_d)
\end{equation}

Where $|W_r|$ represents the number of words in $W_r$ except stop words.
It is important to mention that the stop word removal does not impact the rationality of the proposed method.
However, mutual stop words in the result and the prototype documents do not indicate that both documents are similar semantically and thus discarded.

Note that, although the $WMD$ is a binary function defined on pairs of documents, it is not a distance metric. 
The $WMD$ is not symmetric and $WMD(r,d)=0$ does not mean that $r$ and $d$ are equal in any sense. 
Rather $WMD$ is similar to a fuzzy version of set inclusion (\(\subseteq\)), where $WMD(r,d)=0 \implies W_r\subseteq W_d$.
If $r$ contains only words in $d$ or their synonyms, $WMD(r,d)$ will be close to zero.  
The WMD works best when $r$ is shorter than $d$ since $d$ may be covering multiple topics that are not mentioned in $r$.

In the final step, we set the mean word mover's distance (MMD) measure to estimate the similarity of the multiple results to a given prototype document.  
Let $R$ be a set of short documents retrieved from a search engine.  
We define the $MMD$ as the mean $WMD$ of all results $r\in R$ with respect to the prototype $d$ (see Equation~\ref{eq:mre}):
\begin{equation}
\label{eq:mre}
    MMD(R, d) = \frac{1}{|R|}\sum_{r\in R}WMD(r, d)
\end{equation}

The MMD defined above is designed to measure only one aspect of query performance, the relevance of the results. 
Other important aspects, for example, the number of results, are intentionally not captured by the MMD.
The quality of the MMD is affected by the quality of the underlying word embedding model. 
For general purpose query evaluation, it is recommended to use word embedding models trained on large non-domain specific datasets.

\subsection{Iterative Query Selection}
\label{sec:hill_climbing}

The proposed \emph{iterative query selection (IQS)} method is based on a local search algorithm, which selects the queries that maximize the relevance of the corresponding results retrieved from an opaque search engine.
We use the hill climbing algorithm~\cite{skiena1998algorithm} since querying the search engine is resource-intensive and we need to find local optimum with only a few iterations~\cite{skiena1998algorithm}.

Let $d$ be a prototype document and $W_d$ be the set of words in $d$ as in the previous subsection. 
$W_d$ does not contain stop words. 
In addition, named entities, e.g., ``Michael Jordan,'' are considered as a single term if they are found in the vocabulary of the word embedding approach used as the basis for the WMD.  

Let $V_d$ denote the vocabulary of terms from which possible queries $q\in V_d$ are selected. 
$V_d$ may be equal to $W_d$ or expanded using any query expansion approach. 
We consider two query expansion methods: 
(1) Adding synonyms based on WordNet~\cite{miller95} for each word in $W_d$, later referred to as \emph{Syn}.
(2) Adding $k$ closest words in the embedding space for each candidate word in $W_d$, later referred to as \emph{KNN}.

The IQS searches through the space of possible queries $q\in V_d$. 
It starts with a random subset of words from $V_d$. 
For efficiency, the query size is limited by two control variables $minq$, and $maxq$, which are the minimal and the maximal number of words in a query.  
In every iteration, we randomly modify the query using one of the following three actions: 
$\textsc{AddWord}(q,V_d)$ randomly adds to $q$ a word from $V_d$ that is not yet in $q$.
$\textsc{RemoveWord}(q,V_d)$ removes a random word from the query $q$ decreasing its size.
$\textsc{SwapWords}(q,V_d)$ exchanges a random word in $q$ with a random word in $V_d$ that was not already in $q$. 
Possible actions are chosen to ensure the query size constraints.

After modifying the query $q$ using one of the three actions, we evaluate the MMD of its results $R_q$ from the search engine $se$.
Due to computational and network performance considerations, it is important to limit the number of results retrieved from $se$ in each iteration of the algorithm. 
Usually, this limit further referred to as $rlimit\geq|R_q| $, is defined by the search engine interface and is set to the number of results on a single page. 
The larger the $rlimit$ is, the more accurate the $MMD(R_q,d)$ since it will be calculated on more retrieved documents. 
However, the $rlimit$ is also the primary factor (linearly) affecting the time of an iteration.

The hill climbing IQS algorithm is implemented as described in Algorithm~\ref{alg:hill_climb_search}.
It receives as an input a prototype document $d$, an opaque search engine $se$, the maximal and minimal number of words in a query ($maxq$ and $minq$, respectively), the maximal number of iterations $itr$, and the number of result documents ($rlimit$) retrieved from $se$ in each iteration.
During the algorithm, we keep only queries that decrease the $MMD$ score. 
If the query returns no results, we mark the query as irrelevant by setting its $MMD(R_q,d)$ score to be the maximal score of 2.

The IQS returns an ordered set of queries.
Some search engines allow words from the query to be missing in the results while others retrieve only documents containing all keywords in the query. 
Twitter is an example of the latter, a boolean search engine. 
In the case of a boolean search engine, it is important to run multiple slightly modified queries in order to retrieve as many relevant results as possible. 
This is the main reason due to which the IQS returns a list of queries and not only a single best query. 

\begin{algorithm}
\caption{Iterative Query Selection}
\label{alg:hill_climb_search}
\begin{algorithmic}[1]

\Procedure{BuildQueries}{$d, se, itr, minq, maxq, rlimit$} 
    \State $queries \gets$ empty priority queue
    \State $V_d \gets$ filtered and expanded set of words in $d$ 
    \State $q_{best} \gets$ random subset of $V_d$
    \State $R_{q_{best}} \gets se(q_{best},rlimit)$ 
    \State calculate $MMD(R_{q_{best}}, d)$ 
    \State $q_{new} \gets q_{best}$
    \State $R_{q_{new}} \gets R_{q_{best}}$    
    \Loop{ $itr$ times}
        \State $actions = \{\textsc{AddWord, RemoveWord, SwapWords}\}$
        \If{$|q_{new}| = maxq \vee |R_{q_{new}}|=0$}
            remove $\textsc{AddWord}$ from $actions$
        \ElsIf{$|q| = minq$}
            remove $\textsc{RemoveWord}$ from $actions$ 
        \EndIf
        \State $action \gets random(actions)$
        \State $q_{new} \gets action(q_{best}, V_d)$
        \State $R_{q_{new}} \gets se(q_{new},rlimit)$ 
        \State Using Eq.(3), calculate $MMD(R_{q_{new}}, d)$ 
        \If{$MMD(R_{q_{new}},d) < MMD(R_{q_{best}},d)$}
          \State $queries.add(q_{new},MMD(R_{q_{new}},d))$
          \State $q_{best} \gets q_{new}$
        \EndIf
    \EndLoop
    \Return $queries$ 
\EndProcedure

\Procedure{AddWord}{$q, V_d$}
    \Return $q \cup random(V_d \setminus q)$
\EndProcedure

\Procedure{RemoveWord}{$q, V_d$}
    \Return $q \setminus random(q)$
\EndProcedure

\Procedure{SwapWords}{$q, V_d$}
    \Return $\textsc{RemoveWord}(\textsc{AddWord}(q, V_d),V_d)$
\EndProcedure

\end{algorithmic}
\end{algorithm}


\section{Experiments}
\label{sec:methods_evaluation}
In this section, we describe a series of experiments that evaluate our iterative query selection (IQS) method.
Since IQS required a loss function that evaluates each generated query, we first measure the ability of the WMD (word mover's distance) measure to distinguish between relevant and irrelevant results (see Section~\ref{sec:mre_active_retrival_evaluation}).
Then we examine the WMD performance correlation to the informativeness of the prototype document.
After evaluating the WMD, we examine the performance of the IQS using the MMD (mean WMD) as an active retrieval method for an opaque search engine (see Section~\ref{sec:hill_climbing_evaluation}).

\subsection{Datasets}
In the following experiments, we used the \emph{Twitter TREC Microblog 2012} dataset and the \textit{TREC-COVID 2019} dataset.
The \emph{Twitter TREC Microblog 2012} consists of 59 topics (used as initial queries) and 73K judgments (relevant and irrelevant tweets) for those topics~\cite{soboroff2012overview}.
The corpus was collected over two weeks, from January 23, 2011, to February 7, 2011, containing 16M tweets.
The \textit{TREC-COVID 2019}~\footnote{\url{https://www.kaggle.com/c/trec-covid-information-retrieval/overview}} consists of 35 topics and 20.7K judgments.
It was collected from COVID-19 Open Research Dataset (CORD-19)~\footnote{\url{https://www.semanticscholar.org/cord19}} that contains biomedical articles related to COVID-19.
This dataset was constructed to develop solutions that improve searching for reliable information on the virus and its impact.

\subsection{The Word Mover's Distance (WMD) Evaluation}
\label{sec:mre_active_retrival_evaluation}
The WMD's purpose is to rank documents by their relevance to a prototype document.
However, The Twitter TREC Microblog 2012 dataset and TREC-COVID 2019 dataset include topic definitions that cannot be used as prototype documents (initial query), due to their rather short length. 
Therefore, we iteratively construct such prototype documents for each topic using a relevance feedback process.

\subsubsection{Experimental Setup}
\label{sec:mre_experimental_setup}
The constructed prototype document should reflect the topic being searched. 
We use the following general process to iteratively build a prototype document for each topic and improve the results retrieved using the WMD.
First, we use the initial query for each topic as the prototype document. 
We calculate the WMD between the prototype document and each tweet in the dataset.
Although the prototype document is too short, some relevant tweets can be found using the WMD. 
Second, we retrieve the top $k$ results and request relevance feedback from a user (or from an oracle if ground truth is provided for evaluation purposes). 
Next, we expand the prototype document using the content of the relevant retrieved results and run the second step again.

It is important to note that the relevance feedback should be saved to avoid labeling the same result multiple times.
The process stops after $n$ labeled results for each topic in the dataset (or a user is satisfied with the results).
In this experiment, we set the top $k$ results to 10 and $n$ to be $300$.
For the TREC Microblog 2012 dataset, we discard query 76, since it does not contain any judgments labeled as relevant.

As baselines we use the following:
Okapi BM25~\cite{robertson2009probabilistic}, latent
semantic analysis (LSA)~\cite{deerwester1990indexing}, and TF-IDF.
Also, we use the dual embedding space model (DESM) using the same pre-trained
word embeddings used for the WMD~\cite{nalisnick2016improving}.
In this comparison, we use only unsupervised document similarity measures since our proposed method should run on any search engine without training.
Note that MB25, LSA, and TF-IDF are not purely unsupervised measures since they require knowledge of the corpus which can be argued as training.
Since we have knowledge of the corpus in this experiment, we consider them unsupervised.

\subsubsection{Results \& Discussion}
\label{sec:mre_results}

The mean average precision (MAP) and R-precision results are summarized in Table~\ref{tab:mre_evaluation}.
As can be seen, the WMD outperforms other methods evaluated on both datasets in terms of MAP and R-precision. 
These results emphasize the effectiveness of the WMD in being a good indicator for distinguishing between relevant and irrelevant documents. 
Also, it strengths the pre-trained word vectors of being very useful to detect similar words in two documents. 

\begin{figure}[h!]
\centering
\includegraphics[scale=0.55]{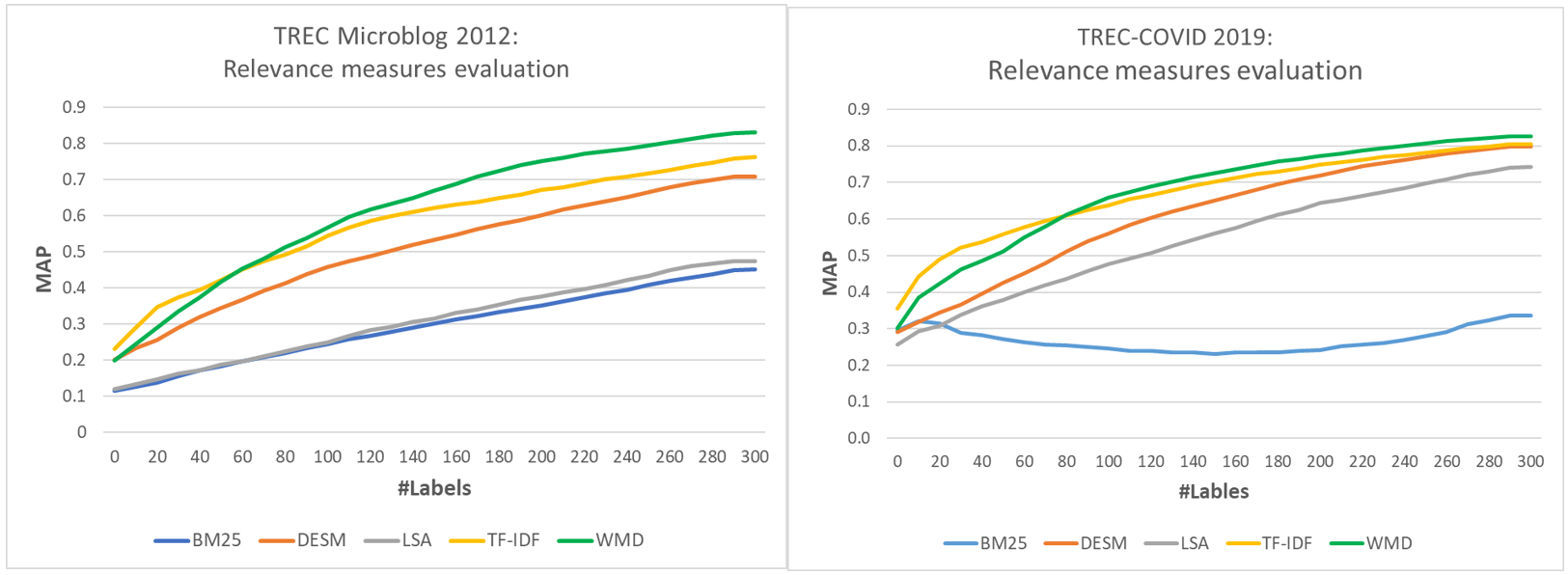}
\caption{Performance evaluation of the relevance measures on the Twitter TREC Microblog 2012 and TREC-COVID 2019.}
\label{fig:elevace_measures_MAP_to_Labels}
\end{figure}

Another impotent aspect we want to examine is the effect of the prototype document informativeness on the MAP score.
This aspect is impotent because it examines whether the relevance measure estimates the results' relevance according to the prototype or not.
The results are presented in Figure~\ref{fig:elevace_measures_MAP_to_Labels}.
The trends in the results show how the WMD utilizes better the information found in the prototype document to rank the results.
This finding indicates that the WMD is the best candidate as a loss function for our query selection method.

\begin{table}[]
\centering
\begin{tabular}{l|l|l|l|l|}
\cline{2-5}
                                      & \multicolumn{2}{l|}{\textbf{TREC Microblog 2012}}                              & \multicolumn{2}{l|}{\textbf{TREC-COVID 2019}}                                  \\ \hline
\multicolumn{1}{|c|}{\textbf{Method}} & \multicolumn{1}{c|}{\textbf{MAP}} & \multicolumn{1}{c|}{\textbf{R\_Precision}} & \multicolumn{1}{c|}{\textbf{MAP}} & \multicolumn{1}{c|}{\textbf{R\_Precision}} \\ \hline
\multicolumn{1}{|l|}{BM25}            & 0.451                             & 0.384                                      & 0.335                             & 0.345                                      \\ \hline
\multicolumn{1}{|l|}{LSA}             & 0.474                             & 0.410                                      & 0.741                             & 0.676                                      \\ \hline
\multicolumn{1}{|l|}{DESM}            & 0.707                             & 0.639                                      & 0.798                             & 0.749                                      \\ \hline
\multicolumn{1}{|l|}{TF-IDF}          & 0.763                             & 0.694                                      & 0.805                             & 0.759                                      \\ \hline
\multicolumn{1}{|l|}{\textbf{WMD}}    & \textbf{0.831}                    & \textbf{0.780}                             & \textbf{0.825}                         & \textbf{0.788}                                  \\ \hline
\end{tabular}
\caption{Evaluation of the relevance measures on the Twitter TREC Microblog 2012 dataset and TREC-COVID 2019 dataset.}
\label{tab:mre_evaluation}
\end{table}


\subsection{Iterative Query Selection with Relevance Feedback}
\label{sec:hill_climbing_evaluation}
In this experiment, we evaluate the full \emph{iterative query selection (IQS)} pipeline based on a process that mimics interaction with Twitter's search engine. 
Twitter uses a boolean retrieval model, which means that the returned results must contain all the words in the query.
We assume an opaque search engine ($se$) and access the corpus through the boolean search process, like in Twitter.

\subsubsection{Experimental Setup}
\label{sec:iqs_experimental_setup}
Similar to the previous experiment, we dynamically construct a prototype document for each topic using relevance feedback.
First, we use the topic definition as the prototype document.
Second, we run the first iteration of the IQS calculating the MMD (mean WMD) score between the prototype document and the results retrieved.
Third, before proceeding to the next iteration of the IQS, we sort the retrieved results in ascending order, according to their MMD score.
Fourth, we take the top $k=10$ results and ask a user (or an oracle) to label them.
Fifth, we expand the prototype document using the content of the relevant results identified by the user (or the oracle) and proceed to the next iteration of IQS.
The stopping condition is the same as in the previous experiment, retrieving labeling $n=300$ tweets. 
We set the minimal and maximal number of words in a query to be between 1 to 6 ($minq=1, maxq=6$).
This parameter influences the number of retrieved results directly. 
When a query containing a single word, many results are expected to be retrieved that probably most of them are not relevant directly to the prototype document. 
For example, assume that the prototype document is "Crude oil production in the U.S." and the query contains the word "oil", solely.
In this case, the Twitter search engine is expected to retrieve many tweets that include the word although they are not related directly to the oil industry in the U.S. (for example, tweets that focus on oil painting and oil production in Russia).
In the same manner, a high number of words decrease the number of retrieved results but most of these tweets are expected to be relevant to the given prototype document.
Lastly, we set the number of returned results to 20 ($rlimit=20$) to simulate a similar number of retrieved documents from a standard search engine within the Web.

In order to choose the best hyper-parameters for IQS, we tested several ranges: $itr$ between 10 to 45, $runs$ between 1 to 3, and $numQueries$ between 5 to 50. 
For our final evaluation, we used the hyper-parameter configuration that yielded the best results on both datasets.
During the hyper-parameter tuning, we limited the total number of interactions with the given search engine ($runs * itr$) to be up to 45, due to time constraints.
Each search interaction with the Twitter search engine takes approximately 1.4 seconds.
Therefore, the total run time for a claim is about a minute.
Eventually, the best parameters found for IQS were $itr=15$, $runs=3$, $minSize=1$, $maxSize=6$, and $numQueries=40$.

\subsubsection{Results \& Discussion}

We compared the performance of the IQS to the ReQ-ReC implementation on GitHub\footnote{https://github.com/lookatmoon/ReQ-ReC-demo} with the same settings: 
the top 10 documents are labeled by the user ($k=10$), the algorithm stops after 300 labels for each topic ($n=300$), and a boolean search engine.

In addition, we run the ALMIK method, a state-of-the-art active retrieval method proposed by Zheng and Sun~\cite{zheng2019collecting}, on the TREC Microblog 2012 and TREC-COVID 2019 datasets.
We implemented the ALMIK method based on the method description presented in their paper.
Again, we limit the number of label requests from the user to 300 and use the same search engine mechanism.
We also conducted hyper-parameter tuning for ALMIK in order to achieve the best results.
The best results were achieved when the ALMIK conduct 3 rounds of active learning phases.
Between the phases, we conduct a keyword expansion phase and used the new results in the next round.
In each active learning phase, we conduct 10 iterations of 10 label requests of the most uncertain tweets from the user (A total of 100 in each active learning phase).

We reported the performance of the IQS, the ReQ-ReC, and the ALMIK all topics in both datasets Table~\ref{tab:black_box_evaluation}.
The proposed IQS method outperformed the ReQ-ReC and ALMIK in terms of MAP and R-Precision in both datasets.
The results show that the IQS can retrieve more relevant results from a boolean opaque search engine using relevance feedback given a short initial query.

\begin{table}[]
\centering
\begin{tabular}{l|l|l|l|l|}
\cline{2-5}
                                      & \multicolumn{2}{l|}{\textbf{TREC Microblog 2012}}                              & \multicolumn{2}{l|}{\textbf{TREC-COVID 2019}}                                  \\ \hline
\multicolumn{1}{|c|}{\textbf{Method}} & \multicolumn{1}{c|}{\textbf{MAP}} & \multicolumn{1}{c|}{\textbf{R\_Precision}} & \multicolumn{1}{c|}{\textbf{MAP}} & \multicolumn{1}{c|}{\textbf{R\_Precision}} \\ \hline
\multicolumn{1}{|l|}{ReQ-ReC}         & 0.147                             & 0.198                                      & 0.002                             & 0.014                                      \\ \hline
\multicolumn{1}{|l|}{ALMIK}           & 0.164                             & 0.172                                      & 0.288                             & 0.336                                      \\ \hline
\multicolumn{1}{|l|}{\textbf{IQS}}    & \textbf{0.357}                    & \textbf{0.356}                             & \textbf{0.508}                    & \textbf{0.507}                             \\

\hline
\end{tabular}
\caption{
Active retrieval method Comparison on the TREC Microblog 2012 and TREC-COVID 2019 datasets.}
\label{tab:black_box_evaluation}
\end{table}

\subsection{Iterative Query Selection with Keyword Expansion}

In the above experiments, we used the \emph{vanilla} IQS without any additional query expansion methods. 
Alongside the vanilla configuration, we also added two keyword expansion methods associated with IQS.
In the first method, we expand each candidate word using its top five synonyms from WordNet (IQS+Syn).
In the second method, each candidate word is expended using its five nearest neighbors words based on \textit{FastText} word embedding (IQS+KNN).

\subsubsection{Experimental Setup}

We evaluated the performance of the three configurations (IQS, IQS+Syn, and IQS+KNN) with respect to the number of iterations, as follows:
First, we executed each configuration 
with 85 iterations to examine where each configuration converges.
In addition, we executed each configuration 3 times and presented the average performance.
After every 5 iterations, we measured the MAP of the 5 queries selected.
There are cases in which the query with the lowest score retrieves irrelevant documents.  
To avoid these cases, we used the top 5 queries to smooth the variation in performance.
Figure~\ref{fig:hill_iteration_test} presents the performance of the configurations where each sub-figure reflects the number of labels retrieved using the active retrieval process.

\subsubsection{Results \& Discussion}



\begin{figure}[]
\centering
\includegraphics[scale=0.45]{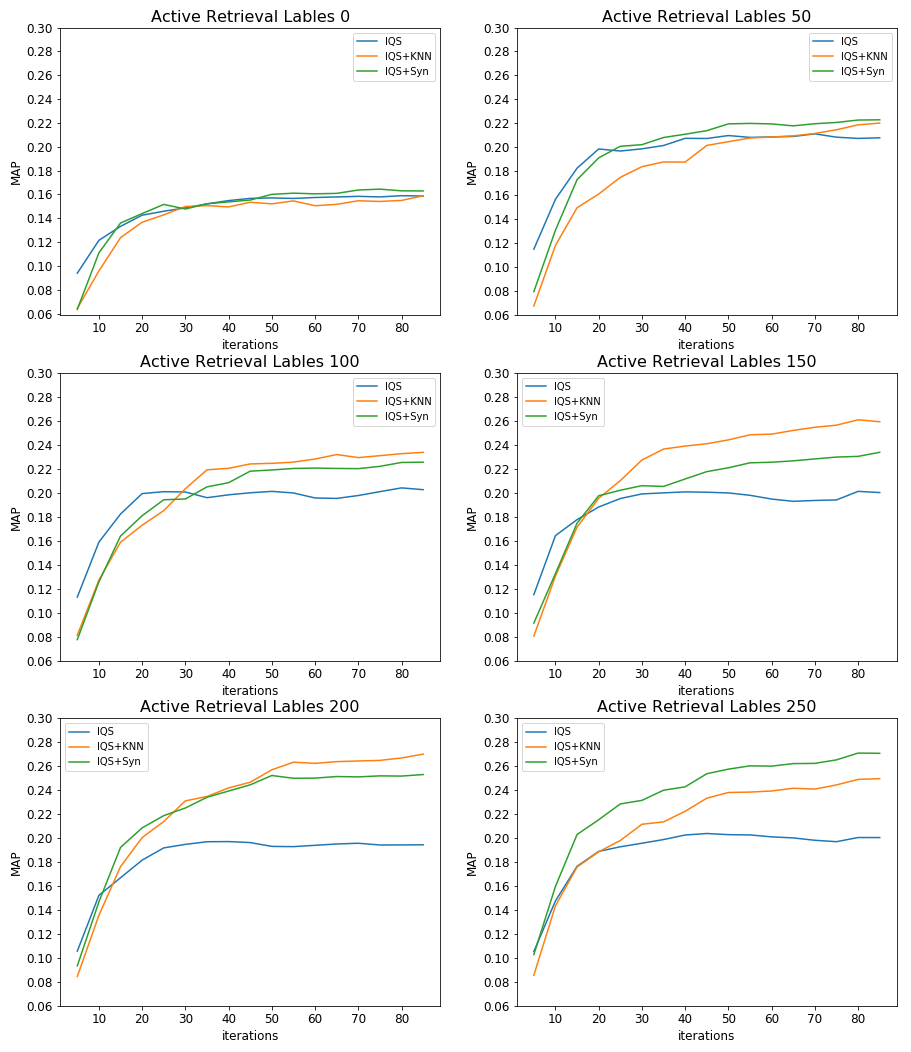}
\caption{Performance evaluation of the IQS configurations on the Twitter TREC Microblog 2012. 
Each graph presents the MAP score of the top five selected queries after every five hill climbing iterations.}
\label{fig:hill_iteration_test}
\end{figure}

As can be seen in Figure~\ref{fig:hill_iteration_test}, vanilla IQS is found superior until a turning point where the IQS+KNN and the IQS+Syn outperformed the vanilla configuration.
In addition, we see that vanilla IQS converged faster than the other configurations.
It is expected since the vanilla IQS has a smaller search space than IQS+KNN and IQS+Syn configurations i.e., the algorithm requires fewer search iterations to converge.
We can also see that the active retrieval iterations affect the performance of the configurations.
The more active retrieval iterations the sooner the turning point arrives.
The reason is that the ratio between the number of original keywords to the number of expended keywords decreases when the prototype is more informative.
A low ratio means that there is a higher likelihood of using more original keywords and thus perform similarly to the vanilla configuration at the start.
Since we want to keep a low number of interactions with the search engine, vanilla IQS is the best option.

\section{Application for Fake News Detection with Pseudo Relevance Feedback}
\label{sec:fake_news_classification}

In many cases, it is not practical and scalable to ask for feedback from the users continuously. 
Therefore, here, we are in a mode of pseudo relevance feedback. 
This means that we apply the \emph{iterative query selection (IQS)} with the mean word mover's distance (MMD) as pseudo relevance feedback.
To reduce the number of interactions with the Twitter search engine we use the IQS vanilla configuration.

In this section, we demonstrate the importance of the IQS as a necessary link in the pipeline of fake news detection. 
Using the proposed IQS and MMD, we demonstrate an automated collection of relevant tweets associated with labeled news items (ground truth). 
Later, using these relevant tweets, we demonstrate fake news detection using supervised machine learning classifiers.
First, we describe the background of fake news detection on online social media (OSM), including the data collection process.
Afterward, we present the dataset obtained using the IQS for fake news detection on OSM.
Finally, we train a classifier on the collected data and present its performance.

\subsection{Data Collection With IQS}
\label{sec:iqs-fake-news-collection}
In this section, we describe the construction of a large dataset for the task of fake news detection on OSM using the \emph{iterative query selection (IQS)}. 
First, we crawl news items from fact-checking websites, such as Snopes, Gossip Cop,\footnote{https://www.gossipcop.com/} and Politifact.
For Snopes and Politifact news items, there are five fine-grained multiple labels: true, mostly true, false, mostly-false and pants-on-fire.
Similar to Rasool et al.~\cite{rasool2019multi}, we converted the classification problem into binary by categorizing the news items that their label are true and mostly-true as true and news items that their label are false, mostly-false and pants-on-fire as false.
For Gossip Cop, we categorized news items with scores between 0 to 3 as false and news items with scores between 7 to 10, as true.
Since the majority of the news items in fact-checking websites are false, we added news items from 10 well-known 
news sources (Time of Israel, CNN News, ABC News, BBC News, The New York Times, The Jerusalem Post, The American Conservative, MSNBC, Fox News, and Politico) as true news items.
A large number of studies exploited reliable news sources as a proxy for true news items~\cite{monti2019fake}.
In total, we collected 70,018 news items (16,212 false, 53,806 true).
For each news item, we set the IQS algorithm to run on Twitter API three times, with the following parameters:
Returning 5 final queries ($numQueries=5$, 
a maximal number of 15 iterations ($itr=15$), five queries,
the number of words in a query is between 3 to 6 ($minq=3, maxq=6$), and the number of results returned equals 20 ($rlimit=20$).
Then, after obtaining the top five final queries from all three runs, we use them in order to retrieve the most relevant tweets for each news item, while limiting the number of tweets returned for each query to 500.
To make the fake news detection more realistic, we collected only the tweets that were posted before the fact-checker assigned a label for the given news item. 
Finally, utilizing this approach, we constructed a large fake news dataset containing about 70,000 news items and about 61 million corresponding posts, the distribution of which is shown in Table~\ref{tab:fake_news_dataset_statistics}.

\begin{table}[h]
\centering
\begin{tabular}{|l|l|l|l|l|l|}
\hline
 
\textbf{Domain}           & \textbf{\#News Items} & \textbf{\#True} & \textbf{\#False} & \textbf{\#Authors} & \textbf{\#Tweets} \\ \hline
politifact.com            & 12,952            & 5,288           & 7,664            & 4,274,688          & 12,525,467        \\ \hline
snopes.com                & 4,682             & 845             & 3,837            & 1,721,162          & 3,617,550         \\ \hline
gossipcop.com             & 4,764             & 53              & 4,711            & 863,906            & 2,302,245         \\ \hline
times of israel         & 21,989            & 21,989          & 0                & 5,401,027          & 20,963,687        \\ \hline
Jerusalem Post        & 3,109             & 3,109           & 0                & 1,469,181          & 3,386,702         \\ \hline
New York Times        & 3,453             & 3,453           & 0                & 1,641,931          & 3,247,247         \\ \hline
abc-news                  & 3,685             & 3,685           & 0                & 1,042,084          & 2,410,832         \\ \hline
cnn-news                  & 3,496             & 3,496           & 0                & 1,381,585          & 2,988,373         \\ \hline
fox-news                  & 3,620             & 3,620           & 0                & 1,115,282          & 2,921,356         \\ \hline
bbc-news                  & 3,687             & 3,687           & 0                & 1,181,490          & 2,469,421         \\ \hline
msnbc-news                & 2,410             & 2,410           & 0                & 731,839            & 1,982,974         \\ \hline
politico                  & 1,640             & 1,640           & 0                & 853,716            & 1,844,441         \\ \hline
American Conservative & 531               & 531             & 0                & 440,086            & 619,336           \\ \hline
\textbf{Total}                     & \textbf{70,018}            & \textbf{53,806}          & \textbf{16,212}           & \textbf{22,117,977}         & \textbf{61,279,631}        \\ \hline
\end{tabular}
\caption{Fake news dataset statistics.}
\label{tab:fake_news_dataset_statistics}
\end{table}

\subsection{Fake News Classification}

The task of fake news classification has been studied intensely in recent years. 
Along with the growth of online news, many non-traditional news sources, such as blogs, have evolved in order to respond to users' ``appetite for information." 
In many cases, however, these sources are operated by amateurs whose reporting is often subjective, misleading, or unreliable~\cite{downie2009reconstruction}.
This "everyone is a journalist" phenomenon~\cite{zeng2018danger}, coupled with the flood of unverified news and the absence of quality control procedures to prevent potential deception, has contributed to an increasing problem of fake news dissemination~\cite{conroy2015automatic}.

The spread of misinformation, propaganda, and fabricated news has potentially harmful effects, even including a significant impact on real-world events~\cite{allcott2017social}.
In recent years, it has weakened public trust in democratic governments and their activities, such as the ``Brexit" referendum and the 2016 U.S. election ~\cite{zhou2019fake}.
World economies are also not immune to the impact of fake news; this was demonstrated when a false claim regarding an injury to President Obama caused the stock markets to plunge (dropping 130 billion dollars)~\cite{rapoza2017can}. 
In recent years, due to the threats to democracy, journalistic integrity, and economies, researchers have been motivated to develop solutions for this serious problem~\cite{zhou2019fake} proposing approaches for the detection of fake news based on natural language processing~\cite{zhou2019fakenlp}, investigating the diffusion of news~\cite{vosoughi2018spread}, etc.

\subsection{Fake News Classification Method}
To classify the news items, we extract author- and post-based features.
For author-based features, we applied aggregation functions on various aspects of author demographics, such as registration age, number of followers, number of followees, number of distributed tweets published by the user, etc.
Post-based features include the aggregations of posts metadata, such as retweet count, text length, the time interval between the oldest and newest post, etc.
For all the features extracted from the post, we removed stop words. 
Also, regarding the post's data, we extracted the following features: sentiment, temporal (post diffusion patterns), LDA (variations on the posts' topics), TF-IDF, and word embedding.
For the latter, we used the Glove Wikipedia pre-trained model with 300 dimensions.
For aggregation functions, we used mean, median, max, min, standard deviation, kurtosis, and skewness functions.


\subsection{Fake News Classification Results}
\label{sec:fake_news_results}
For the classification, we tried many combinations of supervised machine learning algorithms and feature subsets.
All classifiers were trained using 10-fold cross-validation.
Eventually, we averaged the results obtained from all the folds.
We determined that the best performing classifier on the test set was the Random Forest with 100 estimators and a max depth of 10. 
This classifier, with 100 features obtained AUC and accuracy of 0.92 and 0.86, respectively.
\textbf{This result is evidence that an application that detects false news based on the OSM can benefit from training on data collected using IQS. }

Also, we analyzed most of the influential features of the best classifier (see Table~\ref{tab:feature_importance}).
\begin{table}[]
\centering
\begin{tabular}{|l|l|l|}
\hline
 
\textbf{No.} & {\textbf{Feature Name}}    & {\textbf{Gini Importance}} \\ \hline
1 & Number of verified authors                      & 0.162                                     \\ \hline
2 & Max number of posts published by claim's authors  & 0.049                                     \\ \hline
3 & Glove\_wikipedia\_model\_300d max dimension 295 & 0.046                                    \\ \hline
4 & Glove\_wikipedia\_model\_300d min dimension 291 & 0.037                                    \\ \hline
5 & Max favorites count of claim's authors          & 0.036                                     \\ \hline
6 & Max followers count of claim's authors          & 0.033                                    \\ \hline
\end{tabular}
\caption{Features listed by Gini importance.}
\label{tab:feature_importance}
\end{table}
The most important feature was the number of verified authors with a Gini importance of 0.162 (see Table~\ref{tab:feature_importance}).
Comparing the distribution of these authors with respect to fake and true news items, we can see the number of verified authors within true news is 3 times higher than in false news items.
These differences were found statically significant (a p-value of 0.0).
Based on this result, we conclude that verified authors are important actors for fake news detection.
The higher their participation, the higher the reliability of the online discussion.

According to the Gini importance, the second, fifth, and sixth highest influential features were aggregations over the news item's authors.
This strengthens the conclusion of Castillo et al.~\cite{castillo2011information} that author-based features are very relevant for fake news detection within the OSM.

In addition, the third and fourth features are aggregations on the word embedding of the news item's posts.
These features indicate that the fixed-length of vector representations of the words consists of the online discussions that can hold the truthfulness of given news items.

These results show that our proposed algorithm can be utilized for solving real-world problems (e.g., the detection of fake news).
In addition, the machine learning classifiers trained on the collected this large dataset using the IQS obtained impressive results.
This strengthens that our method can be very useful for detecting fake news while retrieving relevant data automatically.

\section{Ethical Considerations}
\label{sec:ethical_considerations}
Collecting information from OSM has raised ethical concerns in recent years. 
To minimize the potential risks of such activities, this study follows
recommendations presented by~\cite{elovici2014ethical}, which deal with the ethical
challenges of OSM and Internet communities.

For this study, we proposed a method, that selects queries for a given prototype document to retrieve the maximal number of relevant documents. 
To evaluate the proposed method, we used the Twitter search engine in order to retrieve tweets associated with the given prototype document.
This service collects tweets published by accounts that agreed to share their information publicly. 

\section{Conclusion \& Future Work}
\label{sec:conclusion}
In this paper, we propose an automated \emph{iterative query selection (IQS)} algorithm for improving information retrieval from opaque search engines. 
This method consists of two components: the \emph{mean word mover's distance (MMD)} which estimates the semantic similarity between the retrieved documents to the given prototype document and the iterative algorithm which selects suitable queries based on the mean WMD (MMD).

We evaluated IQS 
on the \emph{Twitter TREC Microblog 2012} and \textit{TREC-COVID 2019} datasets.
The proposed IQS algorithm was found superior to the other two state-of-the-art methods on both datasets. %
Next, we applied IQS for obtaining a large fake news dataset that later was used for the task of fake news detection.  

As a result, we conclude the following:



First, the WMD score is found to be a successful measure for differentiating between relevant and irrelevant documents concerning a given prototype document (see section~\ref{sec:mre_active_retrival_evaluation}).
This result strengthens previous conclusions of Kusner et al. \cite{kusner2015word} related to WMD being effective for document classification.  

Second, the IQS algorithm obtained the highest performance with respect to two state-of-the-art methods: ReQ-ReC \cite{liu2014automatic} and ALMIK~\cite{chy2019query} and found effective for active retrieval task.

Third, it is recommended to use the proposed IQS as part of an automated fake news detection pipeline.
Using this algorithm, we collected a large-scale fake news dataset consisting of 70K news items, 22M accounts, and 61M tweets, automatically. 
Obtaining an AUC of 0.92 and an accuracy of 0.86 using classic machine learning classifiers emphasizes the quality of the large dataset collected using IQS.



One possible direction for future work could be to demonstrate the proposed approach on different OSM platforms, such as Reddit,\footnote{https://www.reddit.com/} Quora,\footnote{https://www.quora.com/} etc.
Another might compare the effectiveness of a fake news detection system using data collected using a source URL versus data collected using our query selection method.

\section{Availability}
This study is reproducible research. 
Therefore, the Fake News datasets is available.\footnote{\url{https://drive.google.com/drive/folders/1nOGYjGoZHxFwaPm0xci-T7V0a90YW448?usp=sharing}} 
Other datasets for evaluation are available upon request.

\bibliographystyle{unsrt}  
\bibliography{mybibfile}

\end{document}